# Energy Dependence of Cu $L_{2,3}$ Satellites using Synchrotron Excited X-ray Emission Spectroscopy

M. Magnuson, N. Wassdahl, and J. Nordgren

*Department of Physics, Uppsala University, P. O. Box 530, S-751 21 Uppsala, Sweden*

**Abstract**

The $L_{2,3}$ X-ray emission of Cu metal has been measured using monochromatic synchrotron radiation. The self-absorption effect in the spectra is shown to be very small in our experimental geometry. From the quantitative analysis of spectra recorded at different excitation energies, the $L_3/L_2$ emission intensity ratio and the partial Auger-width are extracted. High-energy satellite features on the $L_3$ emission line are separated by a subtraction procedure. The satellite intensity is found to be slowly increasing for excitation energies between the $L_3$, $L_2$ and $L_1$ core-level thresholds due to shake-up and shake-off transitions. As the excitation energy passes the $L_2$ threshold, a step of rapidly increasing satellite intensity of the $L_3$ emission is found due to additional Coster-Kronig processes.

## 1 Introduction

During the last decades, much interest has been focussed on multiple vacancy satellites in Auger spectra of the $3d$ transition metals [1, 2, 3, 4, 5]. In particular, the existence of Auger satellites originating from the decay of $L_{2,3}$ core holes in the presence of additional $3d$ vacancies have been studied. The $3d$ vacancies can be produced either as shake-up excitation or shake-off ionization due to relaxation in connection with the core excitation process or as a result of Coster-Kronig (CK) processes preceeding the Auger decay. The corresponding X-ray decay processes can be studied with X-ray emission (XE) spectroscopy. XE spectroscopy is a powerful technique for studying the electronic structure in terms of the partial density-of-states (pDOS), since the decay from the core-excited state with a localized core-hole can be described according to the dipole selection rules. In XE, the vacancy satellites are traditionally referred to as Wentzel-Druyvesteyn (WD) satellites [6] and typically end up on the high energy side of the main line. Relaxation and screening effects are essential for the evaluation of the intensity and energy shift involved in the satellite transitions which is an important reason for studies of satellites.

XE spectra of Cu metal excited with monochromatic synchrotron radiation were presented in a publication by Wassdahl *et al.* [7] and discussed in terms of shake-up/off and CK processes. The results of this publication were discussed and compared with those of different Cu compounds by Kawai *et al.* [8] It was assumed that the CK contribution was too small to significantly contribute to the high-energy XE satellites and could not be discriminated from the solid state process (charge-transfer from the neighbouring atoms). In a comment by Ohno [9], it was concluded, based on the interpretation of Auger spectra, that the dominant part of the high-energy satellite intensity is indeed due to CK decay. To support their interpretation, Kawai *et al.* [10] then further assumed that the XE spectra of Wassdahl *et al.* [7] had been severely affected by the self-absorption effects and needed to be corrected for this.

As will be shown from our new high-resolution XE measurements of pure Cu metal, the self-absorption effect is indeed very small in the grazing-incidence experimental geometry used. As the excitation energy is increased above the $L_2$ core level, a large enhancement of the $L_3$ satellite intensity is found, indicating $L_2 \rightarrow L_3 M$ CK preceeding the X-ray emission. For excitation energies between the $L$ edges, the satellite intensity is slowly increasing with energy, interpreted as due to additional shake transitions. As will be shown, our results indicate that the origin of the high-energy XE satellite should be interpreted as being due to vacancies produced either by shake-up/shake-off upon primary excitation or, above the $L_2$ excitation threshold, mainly owing to CK decay preceeding the XE process.

## 2 Experimental Details

The experiments were performed at beamline 8.0 at the *Advanced Light Source*, Lawrence Berkeley National Laboratory. The beamline comprises a 99-pole, 5 cm period undulator and a spherical-grating monochromator. The end-station built at Uppsala University includes a rotatable grazing-incidence grating XE spectrometer [11] and a Scienta SES200 photoelectron spectrometer [12]. The base pressure was lower than $2 \times 10^{-10}$ Torr during preparations and measurements. The Cu(100) single crystal sample was of high purity and crystal quality, and cleaned by means of cyclic argon-ion sputtering and annealing to remove surface contaminants.

In order to determine the excitation energies, X-ray absorption (XA) spectra in the threshold region were measured by total electron-yield measurements. In order to normalize the XE spectra, the incident photon current was continuously monitored using a gold mesh in front of the sample. During the XA and XE measurements the resolution of the monochromator of the beamline was 0.3 eV and 0.5 eV, respectively. The sample was oriented so that the photons were incident at an angle of $7^o$ and with the polarization vector parallel to the surface plane. The emitted photons were recorded at an angle near normal to the sample surface, perpendicular to the incoming photons, with a resolution slightly better than 0.8 eV. The grazing-in normal-out setup was chosen to minimize self-absorption effects.

## 3 Results and Discussion

Figure 1 shows $L_{2,3}$ XE spectra of Cu normalized to the integrated photon flux, excited at various energies indicated by the arrows in the XA spectrum shown in the inset. The spectra were measured from excitation energies from 932.5 eV, at the $L_3$ threshold, up to energies as high as 1109.1 eV, above both the $L_2$ ($E_B$=952.3 eV) and $L_1$ ($E_B$=1096.7 eV) thresholds. For excitation energies above the $L_3$ threshold ($E_B$=932.5 eV), a growing satellite tail extending towards higher energies on the high-energy side of the main line is clearly observed in the spectra. It should be noted that we do not observe any elastic emission peak due to recombination, even though our experimental geometry with the polarization vector parallel to the surface plane would enhance the probability for such a process to occur.

Self-absorption is known to affect the shape of XE spectra on the high-energy flank of the main peak, where the satellites occur, since there is an overlap in energy between the XE and XA spectra. The corrected observed XE intensity can be written as; $I = I_o[1 + (\mu_{out}/\mu_{in}) \times tan(\tau)]^{-1}$, where $I_o$ is the unperturbed decay intensity, and $\mu_{in}$ and $\mu_{out}$ are the absorption coefficients for the incident and outgoing radiation, and $\tau$ is the incident angle of the photon beam relative to the sample surface (and between outgoing photon and surface normal). We derived the values of these coefficients at the $L_{2,3}$ thresholds by normalizing the XA spectrum to known atomic values ($\mu_{in}$ = 2.20 and 16.97 $cm^2/g$) well below and above the absorption edges [13]. Based on this procedure the maximum self-absorption at the most severe emission energy (932.5 eV) was found to reduce the emitted intensity between 9 and 10 % for all spectra except for the spectrum excited at 949.4 eV just below the $L_2$ threshold where it was 14 % since the penetration depth is higher at this excitation energy. Since the self-absorption effect is small, the spectra shown in the figures were not corrected for this effect.

We consider the transitions from the core-excited states to the final states and will mainly discuss the following processes:

$$^0 4s^1 \rightarrow 3d^9 4s^1$$
$$4s^1 \rightarrow 3d^8 4s^1$$
$$^0 4s^1 \rightarrow 3d^8 4s^2$$

Process (I) is the main line, (II) an initial state $3d$ vacancy satellite, and (III) a final state satellite due to shake-up in the decay step. The main line (I) has an energy of about 930 eV which is the energy difference between the $2p_{3/2}$ core level binding energy at 932.5 eV and the $3d$ band at about 3 eV. In process (II), the initial state has a mean energy of 944.4 eV with a multiplet spread of 4 eV and the final state has a mean energy of 12.8 eV with a multiplet spread of about 2.8 eV. This gives estimated satellite energies between 928.2 and 935 eV which thus overlap with the main line (I) but is slightly shifted towards the high-energy side [14].

For excitation energies above threshold, the spectra can be treated as produced by an incoherent nonresonant 'two-step' process, with separate excitation and emission steps, and the satellite contribution can therefore be separated from the main line by a subtraction procedure [7, 15]. This is illustrated in Fig. 2, where the satellite difference spectrum ($I_S$, dashed area) has been constructed with the 1088.5 eV spectrum subtracted by the resonantly excited 932.5 eV spectrum ($I_R$) by normalizing the spectra so that the low-energy flanks coincide. The difference spectrum represents the part of the 1088.5 eV spectrum which is mainly produced by excitation via the $L_2$ CK preceeded decay.

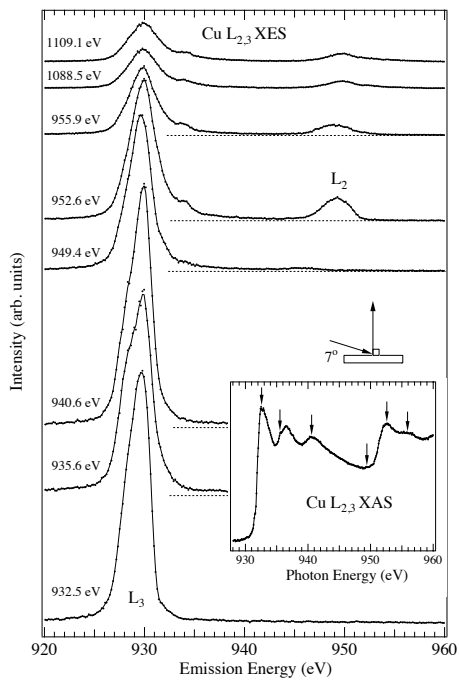

**Figure 1:** A series of XE spectra of Cu measured for excitation energies 932.5 eV to 1109.1 eV. The excitation photon energies are indicated by arrows in the XA spectrum in the inset.

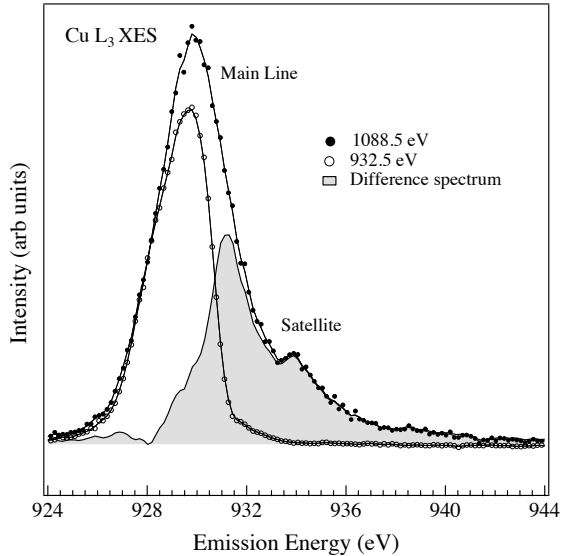

**Figure 2:** Separation of CK preceeded satellite contribution between the spectra excited at 1088.5 and 932.5 eV photon energies. The dashed area represents the difference spectrum.

As can be seen in Fig. 2, the peak position of the XE satellite is shifted by about 1 eV to higher energy and is broader than the main line due to multiplet effects in the initial and final states of the process. Process (III) has the same final state as process (II) but the initial state is solely an $L_3$ core-hole leading to low-energy satellites between 918 and 921 eV. However, although process (III) is possible we do not observe any satellite peaks in the spectra due to this transition.

The direct observation of the X-ray emission satellites makes it possible to obtain the relative inherent widths of the $L_3$ and $L_2$ hole states. Thus, the partial atomic-level widths can be quantitatively extracted by comparing the integrated intensities between the $L_3$ and $L_2$ XE peaks excluding their satellite contributions.

From a statistical point of view, the $2p_{3/2}$ and $2p_{1/2}$ core level ionization ratio is 2:1 for excitation energies between the $L_2$ and $L_1$ thresholds. When using excitation energies above the $L_1$ threshold, the $L_1$ CK decay will also affect the initial core hole population. This may change the subsequent XE $L_3/L_2$ intensity ratio ($I_X^{ratio}$) for the spectrum excited above the $L_1$ threshold. To avoid this problem, we used the XE spectrum excited at 1088.5 eV i.e., the excitation energy is just below the $L_1$ threshold, and the two threshold spectra excited at 932.5 and 952.6 eV to extract the satellite contributions. Including the satellites, $I_X^{ratio}$ was found to be 4.64 using the values given in Table I for the main line ($I_R$) and the satellite ($I_S$) contributions of the $L_3$ and $L_2$ levels. Excluding the satellites, the $I_X^{ratio}$ is found to be 4.48.

Apart from the pDOS spectral distribution, the widths of the XE peaks include Lorentzian lifetime contributions from the core- and final states, also convoluted with the Gaussian instrumental resolution profile. The natural lifetime energy width $\Gamma$ of a state is related to its lifetime $\tau$ by the Heisenberg uncertainty principle, $\Gamma\tau=\hbar$, where h is Planck's constant. We consider the decay from one core hole with characteristic exponential decay-times corresponding to the partial energy widths for the Auger ($\Gamma_A$), radiative ($\Gamma_X$), and Coster-Kronig ($\Gamma_{CK}$) decays, respectively. From this, the $L_3/L_2$ XE intensity ratio can be written as:

$$I_X^{ratio} = \frac{2 \cdot \Gamma_X^{(3/2)}}{\Gamma_X^{(3/2)} + \Gamma_A^{(3/2)}} / \frac{\Gamma_X^{(1/2)}}{\Gamma_X^{(1/2)} + \Gamma_A^{(1/2)} + \Gamma_{CK}^{(1/2)}} \qquad (1)$$

The $L$ shell fluorescence yields are small for Cu, and furthermore the Auger width, XE width, and photoionization cross section are approximately the same at the $L_3$ respectively the $L_2$ thresholds [16]. The X-ray width $\Gamma_X$ is expected to be in the range of a few meV. We try to obtain the Auger width $\Gamma_A$ using experimental values [17, 18] of Cu metal, yielding CK and total widths shown in Table II. The value of $\Gamma_A$ is difficult to obtain explicitly with XPS since the measured total linewidths of the $L_3$ and $L_2$ levels contain contributions from the instrumental broadening, the width of excitation, electron-phonon interactions, many-electron effects etc. $\Gamma_{CK}$ can be determined from the difference in linewidths between the $L_3$ and $L_2$ XPS levels. If we neglect $\Gamma_X$ (since $\Gamma_X \ll \Gamma_A$, $\Gamma_{CK}$) and assume that $\Gamma_A^{(1/2)} = \Gamma_A^{(3/2)}$, the Auger width can be derived from equation (1) as:

$$\Gamma_A = \frac{2 \cdot \Gamma_{CK}^{(1/2)}}{I_X^{ratio} - 2} \qquad (2)$$

We find the Auger decay width $\Gamma_A$, for pure Cu to be between 0.45 and 0.55 eV by using $I_X^{ratio} = 4.48$ and the values of $\Gamma_{CK}$ given in Table II. The theoretical prediction of the Auger width for Cu atoms (gas phase) by Yin *et al.* [1] (0.53 eV) and a many-body Green's function calculation by Ohno *et al.* [19] (0.43 eV) are in good agreement with our experimental results. The extracted $I_X^{ratio}$ value (4.48) is also in fairly good agreement with the $(2p_{3/2} \rightarrow d^2(^1G_4))/(2p_{1/2} \rightarrow d^2(^1G_4))$ Auger main-line intensity ratio of 5.3 obtained by Antonides *et al.* [17].

**Table 1:** Experimental $L_3$ and $L_2$ XE intensities in Cu metal for the main line (R) and satellite (S) contributions at 1088.5 eV excitation energy. The values are normalized to the $L_3$ main line.

| $I_R^{(3/2)}$ | $I_R^{(1/2)}$ | $I_S^{(3/2)}$ | $I_S^{(1/2)}$ |
|---|---|---|---|
| 1.00 | 0.22 | 0.69 | 0.14 |

The difference between the statistical ratio (2:1) and the derived intensity ratio ($\sim$ 4.5/1) indicates that part of the high-energy shoulder of the $L_3$ XE line is related to the $L_2$ ionization through the CK process providing the dominating decay channel for the $L_2$ hole state, leading to a decrease of the fluorescence yield for the $L_2$ state compared to the $L_3$ state. The CK decay from the $L_2$ core level leads to a shorter lifetime and a larger Lorenzian width for this core state than for the $L_3$ state. The CK process furthermore explains the modified shape of the XE spectra as the $L_2$ hole state is excited.

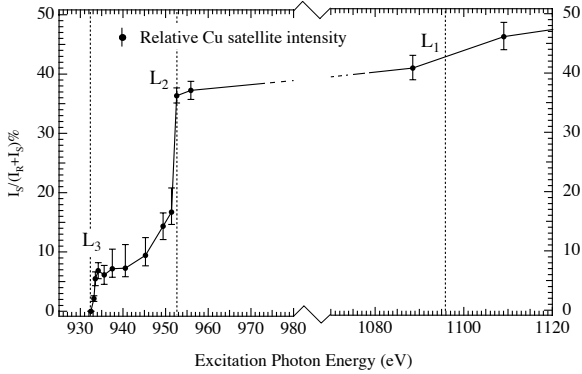

**Figure 3:** The relative XE satellite vs. the main line intensity ratio (in %) shown close to the $L_3$, $L_2$, and $L_1$ thresholds.

Measurements of the relative $L_3$ and $L_2$ XPS intensities and linewidths $\Gamma^{(3/2)}$ and $\Gamma^{(1/2)}$ give an indication of the initial vacancy distribution in the $L_3$ and $L_2$ shells immediately after photoionization. For Cu such XPS experiments indicate that the initial $2p_{3/2}/2p_{1/2}$-shell vacancy ratio is approximately 2:1, whereas $2s$ photoionization is about 6 times weaker than for the $2p_{3/2}$ photoionization [17] at photon energies close above the $2s$ threshold. The $I_X^{ratio}$ value essentially reflects the $L_3/L_2$ vacancy ratio at the time of the X-ray emission. Thus, the large ratio in the spectra indicates that considerable reorganization of vacancies takes place after the excitation of the $L_2$ shell and before the emission of an x-ray photon.

**Table 2:** A comparison between the experimental and calculated values of the Auger width (in eV) obtained from the CK width of: (a) Nyholm *et al.* [18], (b) Antonides *et al.* [17] The calculated values are: (c) atomic by Yin *et al.* [1] and (d) solid by Ohno *et al.* [19].

| $\Gamma_{Tot}$ | $\Gamma_{CK}$ | $\Gamma_A^{exp}$ | $\Gamma_A^{calc}$ |
|---|---|---|---|
| $1.1^{(a)}$ | $0.68^{(a)}$ | 0.55 | $0.53^{(c)}$ |
| $0.90^{(b)}$ | $0.56^{(b)}$ | 0.45 | $0.43^{(d)}$ |

Figure 3 shows the relative $I_S/(I_R+I_S)$ XE satellite intensity ratio (in percent), normalized to the satellite-free $L_3$ threshold spectrum excited at 932.5 eV ($I_R$). The difference spectra have been constructed by normalizing the spectra to the low-energy flanks as described above, using the satellite free threshold spectrum. The error bars of the satellite intensity have been constructed from varying the fit parameters. For excitation energies below the $L_2$ threshold, the satellite has a slowly increasing intensity to less than 10 % of the total intensity, whereas for excitation energies closer to the $L_2$ threshold, a step of very rapid intensity increase is observed up to a new plateau at about 40 %. The satellite intensity is related to the XA cross section which has a peak at the $L_2$ threshold. Above the $L_2$ threshold, the intensity is again slowly increasing due to additional shake processes up towards the $L_1$ level, where additional $L_1$ CK processes are enabled, leading to a new intensity step of about 6 %. This increase of the satellite intensity as the excitation energy passes the $L_1$ threshold is thus much lower than at the $L_2$ threshold. This is mainly due to two things; the difference in photoemission cross section in the excitation process mentioned previously but, also due to the fact that the $L_1$ CK decay is distributed to several decay channels, mainly the $L_1 \rightarrow L_2 M$, $L_1{\rightarrow}L_3M$ and $L_1{\rightarrow}L_3MM$. The intensity steps show that the CK process plays a very important role for the origin of the energy shift and the satellite intensity by the production of extra $3d$ vacancies, preceeding the XE decay.

In the description of the satellite phenomena we focus on the relative importance of excitations in the initial state. When the absorbed photon excites or ionizes an electron and a core hole is created, the surrounding electron cloud will contract in order to screen the positive charge created, causing shake-up excitations or shake-off ionizations in the valence band. The shake intensity depends on the relative configuration coupling due to relaxation. This can be understood with the aid of the Manne-Åberg theorem for inner-shell ionization [20]. The relaxation energy with respect to the unrelaxed Koopmans' energy gives rise to a shake crossection implying WD satellites in the XE spectra.

# 4 Conclusions

X-ray emission spectra of Cu metal have been measured close to the $L_3$, $L_2$, and $L_1$ excitation thresholds with monochromatic synchrotron radiation. Due to the grazing-incidence sample orientation, the spectra are found to be almost free of self-absorption effects. From the quantitative analysis of the spectra, the $L_3/L_2$ intensity ratio and the $L_{2,3}$ Auger-width are extracted. For excitation energies passing the $L_2$ threshold, a sharp step of increasing satellite intensity is found at the $L_3$ emission line, proving the importance of Coster-Kronig decay to the satellite contribution. Thus, we interpret the spectra in the previously accepted picture of initial-state satellites, produced either by shake-up/shake-off during the excitation or as a result of Coster-Kronig decay of inner hole states.

# 5 Acknowledgments

This work was supported by the Swedish Natural Science Research Council (NFR), the Göran Gustavsson Foundation for Research in Natural Sciences and Medicine and the Swedish Institute (SI). ALS is supported by the U.S. Department of Energy, under contract No. DE-AC03-76SF00098. We acknowledge fruitful discussions with S. Butorin and N. Mårtensson and valuable assistance by A. Föhlisch during measurements.

# References

[1]   L. I. Yin, I. Adler, M. H. Chen, and B. Crasemann, Phys. Rev. A **7**, 897 (1973).

[2]   E. D. Roberts, P. Weightman, and C. E. Johnson, J. Phys. C **8**, L301 (1975).

[3]   J. A. D. Matthew, J. D. Nuttall, and T. E. Gallon, J. Phys. C **9**, 883 (1976).

[4]   E. Antonides and G. A. Sawatzky, J. Phys. C **9**, L547 (1976).

[5]   P. Weightman, J. F. McGlip, and C. E. Johnson, J. Phys. C **9**, L585 (1976).

[6]   G. Wentzel, Z. Phys., **31**, 445 (1925), M. J. Druyvesteyn, Z. Phys. **43**, 707 (1927).

[7]   N. Wassdahl, J.-E. Rubensson, G. Bray, P. Glans, P. Bleckert, R. Nyholm, S. Cramm, N. Mårtensson, and J. Nordgren, Phys. Rev. Lett. **64**, 2807 (1990).

[8]   J. Kawai, K. Maeda, K. Nakajima, and Y. Gohshi, Phys. Rev. B **48**, 8560 (1993).

[9]   M. Ohno, Phys. Rev. B **52**, 6127 (1995).

[10]  J. Kawai, K. Maeda, K. Nakajima, and Y. Gohshi, Phys. Rev. B **52**, 6129 (1995).

[11]  J. Nordgren and R. Nyholm, Nucl. Instr. Methods A**246**, 242 (1986); J. Nordgren, G. Bray, S. Cramm, R. Nyholm, J.-E. Rubensson, and N. Wassdahl, Rev. Sci. Instrum. **60**, 1690 (1989).

[12]   N. Mårtensson, P. Baltzer, P. A. Brühwiler, J.-O. Forsell, A. Nilsson, A. Stenborg, and
       B. Wannberg, J. Electr. Spectr. **70**, 117 (1994).
[13]   B. L. Henke, P. Lee, T. J. Tanaka, R. L. Shimabukuro and B. K. Fujikawa,
       At. Data Nucl. Data Tables **27**, 1 (1982).
[14]   I. R. Holton, P. Weightman and P. T. Andrews, J. Phys. C **16**, 3607 (1983).
[15]   N. Wassdahl, PhD thesis, (Uppsala University, Uppsala, Sweden 1987).
[16]   M. O. Krause and J. H. Oliver, J. Phys. Chem. Ref. Data **8**, 329 (1979).
[17]   E. Antonides, E. C. Janse, and G. A. Sawatsky, Phys. Rev. B **15**, 4596 (1977); **15** 1669
       (1977).
[18]   R. Nyholm, N. Mårtensson, A. Lebugle, and U. Axelsson, J. Phys. F **11**, 1727 (1981).
[19]   M. Ohno, J. M. Mariot, and C. F. Hague, J. Electr. Spec. **36**, 17 (1985).
[20]   R. Manne and T. Åberg; Chem Phys. Lett. **7**, 282 (1970).